\documentclass[preprint,prb,eqsecnum,amsmath,amssymb,superscriptaddress]{revtex4}

\usepackage{graphicx}
\usepackage{dcolumn}
\usepackage{bm}

\newcommand{\vecr}{\vec{r}}

\newcommand{\vb}{\vec{b}}

\newcommand{\psiout}{\psi_{\rm out}}
\newcommand{\hatpsiout}{\hat{\psi}_{\rm out}}
\newcommand{\psireact}{\psi_{\rm rc}}
\newcommand{\hatpsireact}{\hat{\psi}_{\rm rc}}

\begin{document}
%
\title{Dispersive diffusion controlled distance dependent recombination  in amorphous semiconductors }
\author{
Kazuhiko Seki}
\email{k-seki@aist.go.jp}
\affiliation{
National Institute of Advanced Industrial Science and Technology (AIST)\\
AIST Tsukuba Central 5, Higashi 1-1-1, Tsukuba, Ibaraki, Japan, 305-8565
}
\author{Mariusz Wojcik}
\affiliation{
Institute of Applied Radiation Chemistry, Technical University of Lodz,
Wroblewskiego 15, 93-590 Lodz, Poland
}
\email{
wojcikm@mitr.p.lodz.pl
}
\author{M. Tachiya
}
\email{m.tachiya@aist.go.jp}
\affiliation{
National Institute of Advanced Industrial Science and Technology (AIST)\\
AIST Tsukuba Central 5, Higashi 1-1-1, Tsukuba, Ibaraki, Japan, 305-8565
}
\begin{abstract}

The photoluminescence in amorphous semiconductors decays according to power law $t^{-\delta}$ at 
long times.
The photoluminescence is controlled by dispersive transport of electrons. 
The latter is usually characterized by 
the power $\alpha$ of the transient current observed 
in the time-of-flight experiments. 
Geminate recombination occurs by radiative tunneling which 
has a distance dependence. 
In this paper, 
we formulate ways to calculate reaction rates and survival probabilities  
in the case carriers execute dispersive diffusion with long-range reactivity. 
The method is applied to obtain tunneling recombination rates under dispersive diffusion.  
The theoretical condition of observing the relation $\delta = \alpha/2 +1$ is obtained and
theoretical recombination rates are compared to the kinetics of observed photoluminescence decay 
in the whole time range measured.  
\end{abstract}

\maketitle

\newpage
\setcounter{equation}{0}
\section{Introduction}
\vspace{0.5cm}

Photoluminescence in amorphous semiconductors exhibits a power law decay 
$\sim t^{-\delta}$ after excitation by weak light pulses at the absorption edge. \cite{Street,NHS,HNS,Oheda,MN85AsS,MN85,Murayama2002,Ando,FDAMurayama,SekiPRB}. 
The power law decay is observed  in the long time range from micro  seconds to milli seconds. 
It is widely recognized that the photoluminescence originates from 
the radiative recombination of electrons and holes in band-tail states \cite{Street,NHS}. 
Holes are quickly trapped and electrons execute hopping-random walks among localized band tail states \cite{Street}. 
At sufficiently high temperatures (T$ >$ 200K), 
the exponent $\delta$ is close to the value $1.5$ 
predicted from the geminate recombination of pairs by normal diffusion. \cite{Street,NHS,HNS,MN85}
However, 
the exponent is smaller than the value $1.5$ at low temperatures. \cite{Oheda,MN85AsS,MN85,Murayama2002,Ando,FDAMurayama,SekiPRB}
Such deviation is interpreted in terms of deviation of motion of charges from the normal diffusion \cite{Oheda,MN85AsS,MN85,Murayama2002,Ando,FDAMurayama,SekiPRB}. 
Murayama and coworkers have systematically investigated the exponent $\delta$ by varying temperature. 
In the experiments, 
the samples of a-Si:H are excited with $10$ nsec light pulses of energy around $2.0$ eV at 
the absorption edge. \cite{MN85} 
The photoluminescence main band is observed at the peak energy $1.22-1.35$ eV and 
the exponent $\delta$ 
 is independent of the excitation intensity in a range from $5 \mbox{[}n\mbox{J}/\mbox{cm}^2 \mbox{]}$ to
$500 \mbox{[} \mu \mbox{J}/\mbox{cm}^2 \mbox{]}$. \cite{MN85}
The measured values of exponent gradually decrease by decreasing temperature. \cite{MN85,Murayama2002,Ando,SekiPRB}
Recently, we have shown that the exponent $\delta$ obeys the relation, 
$\delta=\alpha/2+1$, 
where the disorder parameter $\alpha$ is the exponent characterizing the time evolution of mean square displacements, 
$\left\langle r^2 \left( t \right) \right\rangle \propto t^\alpha$. \cite{SekiPRB} 
$\alpha=1$ corresponds to the normal diffusion and the classical result of $\delta = 1.5$ is reproduced. 
The exponent $\alpha$ is measured by the time-of-flight technique and 
the value decreases linearly with decreasing temperature. \cite{SM,Scher,TiedjePRL,Tiedje,MurayamaMC5,FDAMurayama,SekiPRB}
The transient current measured by the time-of-flight technique is described by 
$I(t) \sim 
t^{-(1-\alpha)}$ when $t< t_T $,
where $\alpha$ is a constant smaller than $1$ 
and $t_T$ is the transit time. 
It is shown based on  a continuous time random walk model that 
$\alpha$ in the transient current is the same as the exponent characterizing the dispersive diffusion, 
$\left\langle r^2 \left( t \right) \right\rangle \propto t^\alpha$. \cite{SM,Scher}
In the time-of-flight experiments, 
electrons move into deeper states while they drift from 
one side of the sample to the other. 
Because electrons in shallow states in the band tails are released more rapidly than those in deep states, 
drift mobility decreases monotonically as time passes. 
The same relation between $\delta$ and $\alpha$ 
is derived from a similar model using 
the percolation approach to the diffusion of charge carriers in disordered systems. \cite{Berlin93}
A simple relation between $\delta$ and $\alpha$, 
$\delta = \alpha/2 +1$, 
is confirmed by experimental data of a-Si:H taken at various temperatures. \cite{SekiPRB}
The relation between $\delta$ and $\alpha$ expressed as $\delta=\alpha/2 +1$ can be 
obtained from a fractional reaction-diffusion  equation which is derived from 
a continuous time random walk model. \cite{SekiPRB} 
In the derivation, 
we have assumed 
that the recombination takes place at a certain distance between 
the electron and the hole of a pair. 
However, 
the most probable mechanism of geminate recombination is radiative tunneling. 
Its rate 
has a distance dependence, described by, 
$k_0 \exp \left( - 2 \beta r \right)$. \cite{Street,NHS,HNS}
In this paper, we generalize our previous results 
to include the effect of the distance dependence of reactivity.

Recently, methods to take account of reaction into fractional diffusion equation have been developed. 
\cite{Yuste,Henry,Sung,Fukunaga,Seki,SekiOctober,FDASeki,Yuste2004,Metzler2004,Shushin,ShushinNJ,Huh}
Interesting features emerged from such studies are 
that reaction terms have a memory effect  
because reaction interferes with random walks consisting of various hopping frequency 
associated with distribution in activation energy. \cite{Seki,SekiOctober,FDASeki}
A fractional reaction-diffusion equation is derived 
from a continuous time random walk model 
by assuming that the reaction takes place at a certain distance. 
In this paper, 
we formulate ways to calculate reaction rates and survival probabilities  
in the case carries execute dispersive diffusion with long-range reactivity. 
In section 2, we derive methods to calculate reaction rates and survival probabilities 
from a continuous time random walk model with long-range reactivity. 
In section 3, we study the square well sink model under dispersive diffusion 
to confirm the validity of our methods by comparing analytical results with those of numerical simulations. 
In section 4, the method is applied to calculate recombination rates in the case of tunneling reactivity 
of electron and hole pairs under dispersive diffusion. 
Section 5 is devoted to discussion and conclusions.

\setcounter{equation}{0}
\section{Distance dependent reaction under dispersive diffusion}
\vspace{0.5cm}

We formulate ways to calculate the survival probability or its decay rate 
in the case long-range reaction proceeds under dispersive transport. 
We consider geminate recombination of a B particle 
starting at $\vec{r}_0$, 
with A.
It is straight forward to generalize results for other initial conditions 
because the model is linear. 
B particle migrates by anomalously slow diffusion, 
$\langle r^2 (t) \rangle \propto t^\alpha$ with $\alpha <1$. 
The anomalously slow diffusion is called dispersive diffusion. 
One of the features for dispersive diffusion is absence of characteristic time of 
jump motion due to trap energy distribution. 
It can be theoretically investigated by a continuous time random walk model 
in which the jumps are performed according to the waiting time distribution. \cite{Montroll65,Kenkre,Weissbook,Haus,Hughes,Balescu,Metzler,Sokolov}
The waiting time distribution of jump motion in the absence of reaction is denoted by 
$\psi \left( t \right)$. 
We consider $\psi \left( t \right)$ 
which is normalized and has an algebraic asymptotic tail, \cite{Tachiya75,Schnorer,Jakobs}
\begin{eqnarray}
\psi (t) = \frac{\alpha \gamma \left( \alpha + 1, \gamma_{\rm r} t\right)}{\gamma_{\rm r}^\alpha t^{\alpha+1}} 
\sim \frac{\alpha \Gamma \left( \alpha + 1\right)}{\gamma_{\rm r}^\alpha t^{\alpha+1}}, 
\mbox{ } \left(t > \frac{1}{\gamma_{\rm r}}\right)
\label{waitingt}
\end{eqnarray}
 where 
$
\gamma (z, p) \equiv \int_0^p e^{-t} t^{z-1} d\,t \mbox{  for } (\mbox{Re} z > 0)
$
is the incomplete Gamma function and $\Gamma (z)$ is the Gamma function. 
\cite{Abramowitz}
If we assume activated release, 
\begin{eqnarray}
\gamma \left( E \right) \equiv \gamma_r \exp \left( - \frac{E}{k_{\rm B} T} \right), 
\label{releaserate}
\end{eqnarray}
and an exponential distribution of activation energy, 
Eq. (\ref{waitingt}) is obtained by calculating, \cite{Scher,Blumen,Schnorer,Jakobs}
\begin{eqnarray}
\psi (t) = \int_0^{\infty} d\,E 
\frac{1}{k_{\rm B} T_0} \exp \left( - \frac{E}{k_{\rm B} T_0} \right) \gamma (E) \exp \left( - \gamma (E) t \right) . 
\label{waiting}
\end{eqnarray}
$\alpha$ in Eq. (\ref{waitingt}) is related to $T$ and $T_0$ in Eq. (\ref{waiting}) through $\alpha=T/T_0$. 
In the absence of reaction 
the above waiting time distribution function leads to dispersive transport, 
$\langle r^2 (t) \rangle \propto t^\alpha$. 
On the other hand, 
the waiting time distribution for the reaction in the absence of random walk is given by, 
\begin{eqnarray}
\gamma_{\rm rc}  \left( \vecr \right) \exp \left[ - \gamma_{\rm rc} \left( \vecr \right) t \right], 
\label{wtdrc}
\end{eqnarray}
where $\gamma_{\rm rc} \left( \vecr \right)$ is the distance dependent reaction rate. 
In the presence of both jump processes and reaction  
the waiting time distribution of making a jump and that of reaction 
are given by \cite{BarzykinPRL,Seki,SekiOctober,SekiPRB}, 
\begin{eqnarray}
\psiout (\vecr, t) &=&  \psi (t) \exp \left( - \gamma_{\rm rc} \left( \vecr \right)  t \right) .
\label{psiout}\\
\psireact \left( \vecr,t \right) &=& \gamma_{\rm rc} \left( \vecr \right) 
\exp \left[ - \gamma_{\rm rc} \left( \vecr \right)  t \right]
\int_t^\infty d\,t_1 \psi(t_1) ,
\label{psireaction}
\end{eqnarray}
respectively.
Eq. (\ref{psireaction}) is the reaction rate multiplied by 
the probability of remaining at the site, 
which decays either by jump motion or reaction. 
Here, 
we list the Laplace transforms of waiting time distribution functions for later use, 
\cite{Barzykin1,Seki,SekiOctober}
\begin{eqnarray}
\hat{\psi} \left(s \right) &=& 1- 
\,_2F_1 \left[1, \alpha, \alpha+1, - \gamma_{\rm r} /s \right]
\sim 1-
\frac{\pi \alpha}{\sin \pi \alpha} \left( \frac{s}{\gamma_{\rm r}} \right)^{\alpha} , 
\label{Laplacepsi}\\
\hatpsiout (\vecr,s) &=& \hat{\psi} \left(s + \gamma_{\rm rc} \left( \vecr \right) \right) 
\sim 1- \frac{\pi \alpha}{\sin \pi \alpha} \left( \frac{s+\gamma_{\rm rc} \left( \vecr \right) }{\gamma_{\rm r}} \right)^{\alpha} , 
\label{Laplacepsiout}\\
\hatpsireact \left( \vecr,s \right)  &=& \frac{\gamma_{\rm rc} \left( \vecr \right) }
{s+\gamma_{\rm rc} \left( \vecr \right) }
\left[ 1 - \hat{\psi} \left(s + \gamma_{\rm rc} \left( \vecr \right)  \right) \right] 
\sim \frac{\gamma_{\rm rc} \left( \vecr \right) }
{s+\gamma_{\rm rc} \left( \vecr\right) }  
\frac{\pi \alpha}{\sin \pi \alpha} \left( \frac{s+\gamma_{\rm rc} \left( \vecr \right) }{\gamma_{\rm r}} \right)^{\alpha} .
\label{Laplacepsireact}
\end{eqnarray}
The equations for the survival probability and the recombination rate can be derived following the method 
presented in a previous paper. \cite{Seki,SekiOctober} 
We first formulate the problem on the basis of  
a discrete model on a periodic lattice of dimension $d$ 
and then take the continuous limit in space. 
We denote the vector characterizing a jump to the nearest neighbor site $j$,  
by $\vb_j$ ($j=1, 2, \cdots, 2d$)
and the jump length by $b$.
The equation for the probability $\eta (\vecr_i,t) \,dt$ of 
just arriving at site $\vecr_i$  in the time interval between $t$ and $t+ d\, t$
is written as,
\begin{eqnarray}
\eta (\vecr_i, t) 
= \frac{1
}{2d}  \sum_{j=1}^{2d}
\int_0^t dt_1  \psiout (\vecr_i-\vb_j,t-t_1) \eta (\vecr_i -\vb_j, t_1) 
+ \delta_{\vecr_i, \vecr_0} .
\label{etadr}
\end{eqnarray}
By subtracting 
$\int_0^t d\, t_1 \psiout (\vecr_i,t-t_1) \eta (\vecr_i, t_1)$ 
from both sides of Eq. (\ref{etadr}),  
we obtain in the small $b$ limit, 
\begin{eqnarray}
\left[1-\hatpsiout (\vecr,s) \right] \hat{\eta} (\vecr,s) =
\frac{b^2}{2} \nabla^2 \hatpsiout(\vecr,s)  \hat{\eta} \left( \vecr, s \right)  +
\delta \left( \vecr - \vecr_0 \right) ,
\label{etacont}
\end{eqnarray}
where the Laplace transform is introduced, 
{\it i.e.}, 
$\hat{\eta}(\vecr,s) = \int_0^\infty d\, t \exp (- st) \eta (\vecr,t)$. 
Eq. (\ref{etacont}) can be rewritten as, 
\begin{eqnarray}
\left[1-\hat{\psi} (s) \right] \hat{\eta} (\vecr,s) =
 \frac{b^2}{2}  \nabla^2 \hatpsiout(\vecr,s)  \hat{\eta} \left( \vecr, s \right) - 
 \delta \hat{\psi}  (\vecr,s) \hat{\eta} \left( \vecr, s \right) +
\delta \left( \vecr - \vecr_0 \right) ,
\label{diffeeqeta}
\end{eqnarray}
where 
$ \delta \hat{\psi}  (\vecr,s) \equiv \psi \left(s \right) -  \hatpsiout  (\vecr,s) $. 
Up to this point 
$b$ has been assumed to be small but finite. 
In the continuous limit, 
we have to take the limit of $b \rightarrow 0$ with 
the generalized diffusion constant given by \cite{Barkai1}
\begin{eqnarray} 
D_\alpha \equiv \frac{\sin \pi \alpha}{2 \pi \alpha} \gamma_{\rm r}^{\alpha} b^2,
\label{defD}
\end{eqnarray}
kept constant. 
Therefore, we should consider the limit, $\gamma_{\rm rc} \left( \vecr \right)/ \gamma_{\rm r} \ll 1$,  
which leads to   
$\hatpsiout \left(\vecr,s\right) \rightarrow 1$ for  $s \rightarrow 0$.  
Since  
$ \hatpsireact \left(\vecr,0 \right) =1-\hatpsiout \left(\vecr,0 \right)$, 
which follows from the fact that particles at a given site perform either 
jump or reaction, {\it i.e.}, 
$\int_0^\infty \psiout \left(\vecr,t\right) d\,t + \int_0^\infty 
\psireact \left(\vecr,t\right) d\, t=1$, 
we find, 
\begin{eqnarray}
\delta \hat{\psi} \left( \vecr, s \right) \rightarrow \hatpsireact  \left( \vecr, 0 \right) \sim 
\frac{\pi \alpha}{\sin \pi \alpha} \left( \frac{\gamma_{\rm rc} \left(\vecr \right)}{\gamma_{\rm r}} \right)^\alpha \mbox{ for } 
s \rightarrow 0 .
\end{eqnarray}
With the aid of the above relation together with Eq. (\ref{Laplacepsi}), 
Eq. (\ref{diffeeqeta}) becomes in the continuous limit as, 
\begin{eqnarray}
s^\alpha \hat{\eta} (\vecr,s) =
D_\alpha \nabla^2 \hat{\eta} \left( \vecr, s \right) - 
k_\alpha \left( \vecr  \right) \hat{\eta} \left( \vecr, s \right) + \frac{\sin \pi \alpha}{\pi \alpha} 
\gamma_{\rm r}^\alpha \delta \left( \vecr - \vecr_0 \right) ,
\label{continuouseta}
\end{eqnarray}
where the generalized reactivity is defined by, 
\begin{eqnarray}
k_\alpha \left( \vecr  \right) \equiv \gamma_{\rm rc}^\alpha \left( \vecr \right) . 
\label{defintrinsicrate}
\end{eqnarray}
The effective reactivity  $k_\alpha \left( \vecr  \right)$ 
depends on $\alpha$ which characterizes the dispersive diffusion. 
The dispersive diffusion competes with  the reaction kinetics. 
As a result, 
the characteristic exponent $\alpha$ enters into the expression of the effective reactivity. 
The recombination rate $R (t)$ satisfies, 
\begin{eqnarray}
R (t) &=& \int_0^t\, dt_1 \int \, d \vecr 
\psireact \left( \vecr, t - t_1 \right) \eta \left( \vecr, t_1 \right) . 
\end{eqnarray}
After the Laplace transform, 
it is expressed in the asymptotic limit of $s \rightarrow 0$ as, 
\begin{eqnarray}
\hat{R} (s) = \int \,d \vecr \hatpsireact \left( \vecr, 0 \right) \hat{\eta} \left( \vecr, s \right) .
\label{rrateL}
\end{eqnarray}
The recombination rate $R(t)$ and the survival probability $N(t)$ are given by, 
\begin{eqnarray}
R(t)&=&  - \frac{\partial}{\partial t} N (t) =\int \,d \vecr \hatpsireact \left( \vecr, 0 \right) \eta \left( \vecr, t \right),
\label{rrate}\\
N(t)&=&  1 - \int \,d \vecr \hatpsireact \left( \vecr, 0 \right) \int_0^t \,dt_1 \eta \left( \vecr, t_1 \right).
\label{subprob}
\end{eqnarray}
Therefore, survival probabilities and recombination rates are expressed in terms of  
the probability $\eta \left(\vecr, t \right) dt$ of just arriving at site $\vecr$ 
in the time interval between $t$ and $t+ d\, t$, 
which is obtained by solution of    
Eq. (\ref{continuouseta}) followed by the inverse Laplace transform.

Although recombination rates and survival probabilities can be calculated by the above method, 
there is more convenient alternative method which naturally reduces to the conventional method 
in the case of normal diffusion, $\alpha = 1$. 
By introducing, 
$\hat{\rho} \left( \vecr, s \right) = \left[ 1-\hat{\psi} (s) \right] \eta \left( \vecr, s \right) /s$, 
Eq. (\ref{continuouseta}) is rewritten as, 
\begin{eqnarray}
s \hat{\rho} \left( \vecr, s \right) - \delta \left( \vecr - \vecr_0 \right)= s^{1-\alpha} 
\left[D_\alpha \nabla^2 \hat{\rho} \left( \vecr, s \right) - 
k_\alpha \left( \vecr  \right) \hat{\rho} \left( \vecr, s \right) 
\right] .
\label{Laplacecontrho}
\end{eqnarray}
After inverse Laplace transform, 
a fractional reaction-diffusion equation is obtained, 
\begin{eqnarray}
\frac{\partial}{\partial t} \rho \left( \vecr, t \right) = \frac{\partial}{\partial t}
\int_0^t d\,t_1 \frac{1}{\Gamma ( \alpha )} \frac{1}{\left( t - t_1 \right)^{1-\alpha}}
\left[ D_{\alpha} \nabla^2 \rho \left( \vecr, t_1 \right)  - k_{\alpha} \left( \vecr \right) 
\rho \left( \vecr, t_1 \right)  \right] .
\label{contrho}
\end{eqnarray}
Eq. (\ref{rrateL}) is transformed into, 
\begin{eqnarray}
\hat{R} (s) = s^{1-\alpha} \int \,d \vecr k_\alpha \left( \vecr \right) \hat{\rho} \left( \vecr, s \right) ,
\label{rraterho}
\end{eqnarray}
by noticing the relation, 
$\hat{\rho} \left( \vecr, s \right) = \left[ 1-\hat{\psi} (s) \right] \eta \left( \vecr, s \right) /s$. 
After the inverse Laplace transform, 
Eq. (\ref{rraterho}) becomes, 
\begin{eqnarray}
R(t)= -\frac{\partial}{\partial t} N(t) =   \frac{\partial}{\partial t}
\int_0^t d\,t_1 \frac{1}{\Gamma ( \alpha )} \frac{1}{\left( t - t_1 \right)^{1-\alpha}} 
k_{\alpha} \left( \vecr \right) 
\rho \left( \vecr, t_1 \right) . 
\label{raterho}
\end{eqnarray}
In principle, the density $\rho_{\rm d} \left( \vecr, t \right)$ should be given in terms of
the probability of remaining at a site,
$\phi  \left( \vecr, t \right) \equiv \int_t^\infty \,dt_1 \left[ \psireact  \left( \vecr, t_1 \right) +
\psiout  \left( \vecr, t_1 \right) \right]$, as 
$\rho_{\rm d} \left( \vecr, t \right) = \int_0^t d\, t_1 \phi \left(\vecr, t -t_1\right) \eta \left( \vecr
  ,t_1 \right)$. 
After the Laplace transform, we obtain
$\hat{\rho}_{\rm d} \left( \vecr, s \right) =  \left[ 1-\hatpsiout  \left( \vecr, s \right)  -\hatpsireact  \left( \vecr, s \right) \right] \eta \left( \vecr, s \right)/s $. 
The Laplace transform of the density is expressed in terms of 
$\hat{\rho} \left( \vecr, s \right)$ as
$\hat{\rho}_{\rm d} \left( \vecr, s \right) =  \left[ 1-\hatpsiout  \left( \vecr, s \right)  -\hatpsireact  \left( \vecr, s \right) \right] \hat{\rho} \left( \vecr, s \right) /\left[ 1-\hat{\psi} (s) \right]  $.
 In the case of normal diffusion, 
 $\alpha = 1$ holds and the above expression for the density reduces to $\hat{\rho} \left( \vecr, s \right)$ 
as shown in  Appendix \ref{appA} 
but they are not equal for $\alpha < 1$. 
Eq. (\ref{raterho})  is derived from Eq. (\ref{rrateL}). 
On the other hand, 
if we assume  $N(t)= \int \, d \vecr \rho \left( \vecr, t \right)$, 
 integration of 
Eq. (\ref{contrho}) over $\vecr$ 
also yields Eq. (\ref{raterho}). 
Therefore, 
the relation
$N(t)= \int \, d \vecr \rho \left( \vecr, t \right)$ should hold, 
although 
$\rho \left( \vecr, t \right)$ without integration over $\vecr$ cannot be interpreted as the real density for $\alpha < 1$.  
Eq. (\ref{raterho}) together with Eq. (\ref{contrho}) is an alternative method to calculate recombination rates and 
survival probabilities.  
They are useful because recombination rates and survival probabilities under dispersive diffusion 
are derived from those under normal diffusion with $\alpha=1$ by substitution, 
$D_1 \rightarrow s^{1-\alpha} D_\alpha$ and $k_1 \left(\vecr \right) \rightarrow s^{1-\alpha} k_\alpha \left(\vecr \right)$
after Laplace transform.  
In the following section, 
we study the validity of such approach by comparing with numerical simulations. 
It should be also noted that the effective reaction rates depend on $\alpha$ 
through Eq. (\ref{defintrinsicrate}).  
In the case of tunneling with rate given by 
 $\gamma_{\rm rc} \left( \vecr \right) = k_0 e^{-2 \beta r}$, 
the effective reactivity $k_\alpha \left( \vecr \right) $ is given by 
$k_\alpha \left( \vecr \right) = k_0^\alpha e^{-2 \alpha \beta r}$.  
The effective tunneling distance $1/\left( \alpha \beta \right)$  increases 
when $\alpha$ is decreased by energetic disorder. 
In the case of energy transfer, $\gamma_{\rm rc} \left( \vecr \right) = k_0 (R_F/r)^6$,  
with F\"{o}rster radius $R_F$, 
the effective reactivity is given by 
$k_\alpha \left( \vecr \right) = k_0^\alpha (R_F/r)^{6\alpha}$. 
The exponent $6\alpha$ is decreased with decreasing $\alpha$ 
and the long-range nature of energy transfer is enhanced. 
Shushin obtained similar results. 
He showed that the effective reaction radius appearing in the steady state 
solutions of Smoluchowski-type stochastic Liouville equations 
 increases with decreasing $\alpha$  for tunneling reaction. \cite{Shushin}
He also showed that 
the kinetics changes at $\alpha = 1/2$ for  energy transfer of F\"{o}rster type. \cite{Shushin}
We do not further investigate the energy transfer  but focus our attention on  
tunneling recombination reactions.

\setcounter{equation}{0}
\section{\label{sec:sim}Square well sink model under dispersive diffusion}
\vspace{0.5cm}

In this section, we consider 
the model in which reactivity constant is $k_0$ within the sphere of radius R. 
\cite{Teramoto,Wilemski,Doi,Tachiya}
In the absence of diffusion, 
the waiting time distribution of reaction is given by Eq. (\ref{wtdrc}) with 
\begin{eqnarray}
\gamma_{\rm rc} \left( r \right) = k_0*S(r) , 
\label{gammarcWF}
\end{eqnarray}
where the reaction function $S(r)$ is given by, 
\begin{eqnarray}
S \left( r \right) = \left\{ 
\begin{array}{lr} 
1 & \mbox{ for } r \leq R  \\
0 & \mbox{ for } r > R . \nonumber
\end{array}
\right. 
\end{eqnarray}
Reaction takes place with the uniform reactivity whenever the distance between reactants 
is less than $R$. 
This model was already studied in the case of 
the normal diffusion with $\alpha=1$. \cite{Teramoto,Wilemski,Doi,Tachiya}
The effective reactivity is obtained from Eq. (\ref{defintrinsicrate}) as 
\begin{eqnarray}
k_\alpha \left( r \right) = k_0^\alpha *S(r). 
\end{eqnarray}
The effective reactivity is scaled by the exponent $\alpha < 1$ under the dispersive transport. 
As described in the last part of the previous section, 
we apply substitution, 
$D_1 \rightarrow s^{1-\alpha} D_\alpha$ and $k_1 \left(\vecr \right) \rightarrow s^{1-\alpha} k_\alpha \left(\vecr \right)$
to the results derived for normal diffusion in the Laplace domain. 
We consider the geminate recombination in which the initial distance between the pair is $r_0$. 
The escape probability, $\varphi \left( r_0 \right) = N \left( \infty \right)$ is obtained by substituting 
$k_0 /D_1 \rightarrow  k_0^\alpha /D_\alpha$ into 
Eq. (5.13) of Tachiya's result, \cite{Tachiya} 
\begin{eqnarray}
\varphi \left( r_0 \right) = 1 - \frac{R}{r_0} \left( 1 - 
\frac{\tanh \left( R \sqrt{k_0^\alpha /D_\alpha} \right)}{R \sqrt{k_0^\alpha /D_\alpha} } \right) .
\label{escWF}
\end{eqnarray}
The Laplace transform of the survival probability is obtained from 
Eq. (5.10) of Tachiya's result, \cite{Tachiya}
\begin{eqnarray}
\hat{N} \left( s \right) = \frac{1}{s} 
\left[1- 
\left( \frac{k_0}{s} \right)^\alpha 
\frac{e^{- (r_0-R) \sqrt{\frac{s^\alpha}{D_\alpha}}}}{\displaystyle 1 + \left( \frac{k_0}{s} \right)^\alpha} \;  
\frac{\displaystyle R \sqrt{\frac{s^\alpha+k_0^\alpha}{D_\alpha}} 
\cosh \left( R \sqrt{\frac{s^\alpha + k_0^\alpha}{D_\alpha}} \right)
- \sinh \left( R \sqrt{\frac{s^\alpha + k_0^\alpha}{D_\alpha}} \right)}
{r_0 \sqrt{\frac{s^\alpha+k_0^\alpha}{D_\alpha}} 
\cosh \left( R \sqrt{\frac{s^\alpha + k_0^\alpha}{D_\alpha}} \right)
+ r_0 \sqrt{\frac{s^\alpha}{D_\alpha}} 
\sinh \left( R \sqrt{\frac{s^\alpha + k_0^\alpha}{D_\alpha}} \right)}
\right] .
\nonumber \\
\label{suvWF}
\end{eqnarray}
The numerical Laplace inversion of Eq. (\ref{suvWF})
by Stehfest algorithm \cite{Stehfest} is employed to get the theoretical results.
We compare these results to those of numerical simulations.

The simulation procedure is the same as that used previously. \cite{Seki,SekiOctober} 
We explain it briefly here. 
In our model
one reactant is fixed at the origin, 
and the other performs a random walk starting from an initial separation, $r_0=2R$. 
The random walk is realized as a sequence of 
detrapping and instantaneous trapping events,  
with the trap energies generated according to the exponential distribution characterized by 
an attenuation parameter $k_{\rm B} T_0$. 
The rate of release from the traps is assumed to be given by 
Eq. (\ref{releaserate}), 
and the detrapping time from a trap of energy $E$  
is obtained from the exponential distribution with the mean value 
$1/\gamma (E) $ . 
The Gaussian distribution of the jump length is assumed 
for each direction of the Cartesian coordinates 
with the standard deviation of $b$ and the zero mean value.
We assume that the reactions occur within a spherical shell 
of radius  $R$  
and the reactivity is given by $k_0$. 
In the simulation, the reactions are modeled in the following way. 
When a reactant is in the reactive sphere, 
we generate the reaction time $t_{\rm rc}$  
from the exponential distribution with the mean value $1/k_0$ , 
and compare it with the detrapping time $t$  
calculated for the current trap. 
If  $t_{\rm rc}<t$ then the reaction occurs, 
otherwise the particle jumps to another trap 
and the simulation continues. 
The trajectory is calculated 
until the particle either reacts or escapes to a large distance $r_{\rm max}=100R$ . 
The simulation is repeated for at least $10^4$ independent trajectories.

The calculated diffusion coefficient is in good agreement with Eq. (\ref{defD}). 
The simulation results of the escape probability against the reactivity are shown in Fig. 1. 
Excellent agreement is found between
the analytical result of Eq. (\ref{escWF}) and simulations
regardless of $\alpha$ values.
In Fig. 2, 
the survival probabilities obtained from the inverse Laplace transform of Eq. (\ref{suvWF}) 
are compared with the results of numerical
simulations for $\alpha = 0.5$. 
Theoretical results 
coincide with the simulation data for the chosen values of reactivity. 
These results confirm the validity of procedure 
to calculate the survival probability and escape probability 
in sub-diffusive media with long-range reactivity 
described 
in the previous section. 
In the next section, 
this procedure is applied to the case of the tunneling recombination  
under dispersive diffusion.

\setcounter{equation}{0}
\section{Tunneling recombination under dispersive diffusion}
\vspace{0.5cm}
\label{section:TD}

In this section, we examine 
recombination rates under the radiative tunneling described by the first order rate, 
$k_0 e^{-2 \beta r}$, 
where $\beta$ is known to be $\sim 0.05-0.1 \mbox{\AA}^{-1}$. \cite{NHS} 
The effective reactivity under the dispersive transport is given by, 
$k_\alpha \left( \vecr \right)= k_0^\alpha \exp \left( - 2 \alpha \beta r \right)$, 
and we have to solve, 
\begin{eqnarray}
s \hat{\rho} \left( r, s \right) - \frac{\delta \left( r - r_0 \right)}{4 \pi r_0^2}= s^{1-\alpha} 
\left[D_\alpha \nabla^2 \hat{\rho} \left( r, s \right) - 
k_\alpha \left( r  \right) \hat{\rho} \left( r, s \right) 
\right] .
\label{tunneringdiff}
\end{eqnarray}
The kinetics of geminate recombination by normal diffusion and tunneling was 
investigated previously. 
However, a perfectly absorbing boundary condition 
was  imposed on 
a sphere of atomic dimensions at the origin with a certain radius smaller than 
the initial distance of a pair of an electron and a hole, $r_0$. \cite{NHS,HNS}
In this paper, we assume recombination occurs only by tunneling 
and impose a perfectly reflecting boundary condition at the radius, $R$.  
By introducing 
 transformation, 
$\hat{\rho} \left(r, s \right) = \hat{\sigma} \left(r, s \right)/r$
,
and a new variable, 
$
y=
\sqrt{k_\alpha/D_\alpha} \exp \left(- \alpha \beta r \right)/\left( \alpha \beta \right)$, 
the homogeneous part of Eq. (\ref{tunneringdiff}) becomes 
the modified Bessel differential equation. \cite{NHS,HNS,Fabrikant,Berlin,Sipp,Burshtein}
Thus, the homogeneous solutions of Eq. (\ref{tunneringdiff}) are, 
\begin{eqnarray}
\hat{p} (r, s) &=& \frac{1}{r} I_{\frac{1}{\alpha \beta}\sqrt{\frac{s^\alpha}{D_\alpha}}} 
\left( \frac{1}{\alpha \beta} \sqrt{\frac{k_\alpha}{D_\alpha}} e^{- \alpha \beta r} \right), \\
\hat{q} (r, s) &=& \frac{1}{r} K_{\frac{1}{\alpha \beta}\sqrt{\frac{s^\alpha}{D_\alpha}}} 
\left( \frac{1}{\alpha \beta} \sqrt{\frac{k_\alpha}{D_\alpha}} e^{- \alpha  \beta r} \right) . 
\end{eqnarray}
General solutions are written by introducing constants, $C_1$, $C_2$ and $C_3$, which are 
determined to satisfy the boundary conditions, 
\begin{eqnarray}
\hat{\rho} \left(r, s \right) = 
\left\{ \begin{array}{lr}
C_1 \hat{p} \left( r, s \right) & r \geq r_0 \nonumber \\
C_2 \hat{q} \left( r, s \right) + C_3 \hat{p} \left(r, s \right) &  r <  r_0 \nonumber
\end{array} .
\right. 
\end{eqnarray}
Solution should be continuous at $r=r_0$ and satisfy Eq. (\ref{tunneringdiff}), 
\begin{eqnarray}
C_1  \hat{p} \left(r_0, s \right) - C_2 \hat{q} \left(r_0, s \right) - C_3 \hat{p} \left(r_0, s \right)
&=&  0 
\label{condition1}
\\
C_2 \frac{\partial}{\partial r_0} \hat{q} \left(r_0 , s \right) + 
C_3 \frac{\partial}{\partial r_0} \hat{p} \left(r_0 , s \right) -
C_1 \frac{\partial}{\partial r_0} \hat{p} \left(r_0 , s \right) &=& 
\frac{s^{\alpha-1}}{4 \pi D_\alpha r_0^2} .
\label{condition2}
\end{eqnarray}
Perfectly reflecting boundary condition at $r=R < r_0$ leads to, 
\begin{eqnarray} 
C_2 \frac{\partial}{\partial R} \hat{q} \left( R, s \right) + 
C_3 \frac{\partial}{\partial R} \hat{p} \left( R, s \right) =0 .
\label{condition3}
\end{eqnarray}
Later we take the limit of $R \rightarrow 0$. 
We have 3 equations for 3 unknown coefficients, $C_1$, $C_2$ and $C_3$. 
By solving Eqs. (\ref{condition1})-(\ref{condition3}), 
the solution in the Laplace domain is found, 
\begin{eqnarray}
\hat{\rho} \left( r, s \right) = \left\{
\begin{array}{lr}
\displaystyle \frac{s^{\alpha-1}}{4\pi D_\alpha \alpha \beta} 
\left(  \hat{q} \left( r_0, s \right) - \frac{ \hat{q} \,\prime \left( R, s \right)}{ \hat{p} \,\prime \left( R, s \right)}  \hat{p} \left( r_0, s \right) \right) 
 \hat{p} \left( r, s \right) & r \geq r_0 , \\
\displaystyle \frac{s^{\alpha-1}}{4\pi D_\alpha \alpha \beta} 
\left(  \hat{q} \left( r, s \right) - \frac{ \hat{q} \,\prime \left( R, s \right)}{ \hat{p} \,\prime \left( R, s \right)}  \hat{p} \left( r, s \right) \right) 
 \hat{p} \left( r_0, s \right) & r< r_0 ,
\end{array}
\right.
\label{lalpacetd}
\end{eqnarray}
where 
$\hat{p} \,\prime (r,s) \equiv \frac{\partial}{\partial r} \hat{p}(r,s)$ and
$\hat{q} \,\prime (r,s) \equiv \frac{\partial}{\partial r} \hat{q}(r,s)$. 
In the above derivation, 
the relation, 
$\hat{p} (r,s) \hat{q} \,\prime (r,s) -\hat{p} \,\prime (r,s) \hat{q} (r,s)  
= \alpha \beta /r^2$, 
is introduced. 
Because the sample is excited with relatively weak light pulses at the absorption edge, 
an electron is created in the vicinity of a hole.  
The solution is simplified when $r_0=R$, 
\begin{eqnarray}
\hat{\rho} \left( r, s \right) = - \frac{s^{\alpha -1}}{4 \pi D_\alpha R^2} 
\frac{\hat{p} \left(r, s \right)}{\hat{p} \,\prime \left( R, s \right)} , 
\end{eqnarray} 
which is further simplified in the limit of $ R \rightarrow 0$ as, 
\begin{eqnarray}
\hat{\rho} \left( r, s \right) =  \frac{s^{\alpha-1}}{4 \pi D_\alpha r }
\frac{I_{\frac{1}{\alpha \beta}\sqrt{\frac{s^\alpha}{D_\alpha}}} 
\left( \frac{1}{\alpha \beta} \sqrt{\frac{k_0^\alpha}{D_\alpha}} e^{- \alpha \beta r} \right)}
{I_{\frac{1}{\alpha \beta}\sqrt{\frac{s^\alpha}{D_\alpha}}} 
\left( \frac{1}{\alpha \beta} \sqrt{\frac{k_0^\alpha}{D_\alpha}} \right)} .
\end{eqnarray}
Laplace transform of the survival probability is calculated by, 
$
\displaystyle 
\hat{N} \left(s \right) = 4 \pi \int_0^\infty r^2 \hat{\rho} \left( r, s \right) d\, r , 
$
and the result is expressed as, 
\begin{eqnarray}
\hat{N} \left( s \right) &=& \frac{s^{\alpha -1}}{D_\alpha \alpha^2 \beta^2} 
\frac{\displaystyle \left(\frac{1}{2\alpha \beta} \sqrt{\frac{k_0^\alpha}{D_\alpha}}
\right)^{\frac{1}{\alpha \beta} \sqrt{\frac{s^\alpha}{D_\alpha}}}}
{\displaystyle  I_{\frac{1}{\alpha \beta}\sqrt{\frac{s^\alpha}{D_\alpha}}} 
\left( \frac{1}{\alpha \beta} \sqrt{\frac{k_0^\alpha}{D_\alpha}} \right)}
\displaystyle \sum_{n=0}^\infty 
\frac{\left( \frac{1}{2\alpha \beta} \sqrt{\frac{k_0^\alpha}{D_\alpha}}\right)^{2n}}
{\left( 2 n + \frac{1}{\alpha \beta} \sqrt{\frac{s^\alpha}{D_\alpha}} \right)^2 
n! \Gamma 
\left(n + 1 +\frac{1}{\alpha \beta} \sqrt{\frac{s^\alpha}{D_\alpha}} \right)} ,
\label{seriesrate}
\end{eqnarray}
where the series expansion of the modified Bessel function is introduced, 
\cite{Abramowitz} 
\begin{eqnarray}
I_\nu \left( z \right) = \left( \frac{z}{2} \right)^\nu \sum_{n=0}^\infty 
\frac{\displaystyle \left( \frac{z}{2} \right)^{2n}}
{\displaystyle n! \Gamma \left( \nu + n+ 1 \right)} .
\label{seriesMdB}
\end{eqnarray}
By taking the limit of $s \rightarrow 0$, 
the escape probability is obtained as, 
\begin{eqnarray}
\varphi \left(r_0=R \rightarrow 0 \right) =\frac{1}{\displaystyle  I_0
\left( \frac{1}{\alpha \beta} \sqrt{\frac{k_0^\alpha}{D_\alpha}} \right)} .
\end{eqnarray}
In the case the intrinsic recombination rate is small, 
$\frac{1}{2 \alpha \beta} \sqrt{\frac{k_0^\alpha}{D_\alpha}} \leq 1$, 
the recombination rate in the long time limit becomes,  
\begin{eqnarray}
R(t) &=& - \frac{\partial N(t)}{\partial t} =
\frac{\alpha k_0^\alpha }
{\displaystyle  \Gamma \left( 1 - \frac{\alpha}{2} \right)
\left(2 \alpha \beta \sqrt{D_\alpha} \right)^3
\left( 1 + \frac{k_0^\alpha}{4 D_\alpha \alpha^2 \beta^2} \right)^2} 
\frac{1}{t^{\alpha/2+1}} ,
\label{recombrateasym}
\end{eqnarray}
as described in Appendix \ref{appB}.
Clearly, 
the relation, 
$\delta = 1 + \frac{\alpha}{2},
$
holds.

In the limit of 
$\frac{1}{2 \alpha \beta} \sqrt{\frac{k_0^\alpha}{D_\alpha}} \ll 1$, 
the recombination rate in the long time limit is simplified as,  
\begin{eqnarray}
R(t) \sim \left[ \frac{\alpha k_0^\alpha}{\Gamma \left( 1 - \frac{\alpha}{2} \right)
\left(2 \alpha \beta \sqrt{D_\alpha} \right)^3} \right]
\frac{1}{ t^{\alpha/2+1}} . 
\label{powerPL}
\end{eqnarray} 
Therefore, the recombination rate depends on the disorder parameter $\alpha$ 
in the form, 
\begin{equation}
R (t) \sim \frac{1}{C^\alpha} \frac{1}{t^{\alpha/2+1}}, 
\end{equation}
where $C = \gamma_{\rm r}^{3/2} /k_0$ is a constant. 
The relation, 
$\delta = 1 + \frac{\alpha}{2},
$
is observable when 
$k_0$ is small compared to the rate of escape of an electron  
from the region 
within the distance $1/(\alpha \beta) \sim 10-30 \mbox{\AA}$ from a hole 
by hopping. 
In this case 
an electron may enter that region 
many times before recombination. 
In the previous paper, 
the measured photoluminescence is shown to obey a scaling relation, 
$
R \left( t \right)  \sim \frac{1}{\displaystyle C^\alpha t^{\alpha/2+1}}  
$, 
where $C^2 \sim 6.8 \times 10^{13} \mbox{[s}^{-1}\mbox{]}$. \cite{SekiPRB}
This value of $C^2$ is obtained from 
the temperature dependence of photoluminescence intensity observed at a delay time of 
$10\mbox{[}\mu\mbox{s]}$. \cite{SekiPRB}
The temperature dependence is assumed to arise from the disorder parameter, 
which is phenomenologically described by a linear function, 
\begin{eqnarray}
\alpha = \alpha_0 + \frac{T}{T_0} , 
\label{alphaT}
\end{eqnarray}
where $\alpha_0 = 0.15$ and $T_0 = 313.4$ [K]. \cite{MurayamaMC5,FDAMurayama,SekiPRB} 
If we assume an exponential distribution of band tail states characterized by an attenuation parameter $k_{\rm B} T_0$, 
$\alpha$ is given by $\alpha=T/T_0$.
The presence of the constant term, $\alpha_0$, in Eq. (\ref{alphaT}) 
implies 
that activated release of an electron from 
an exponentially distributed band tail states is an oversimplified model. 
Nevertheless,  
the linear temperature dependence is a good approximation for 
the observed values of 
the disorder parameter, $\alpha$. 
Eq. (\ref{powerPL}) also has the scaling form with a constant, 
 $C=\gamma_{\rm r}^{3/2} /k_0$. 
Although radiative tunneling rate has a distance dependence, 
photoluminescence can be modeled by recombination reaction 
at encounter distance as we have done in the previous paper, 
as far as long time asymptote is concerned. 
The tunneling rate competes with the jump frequency of electrons only at localized sites in the vicinity of a hole.  
The specific condition is given by $\frac{1}{2 \alpha \beta} \sqrt{\frac{k_0^\alpha}{D_\alpha}} \leq 1$. 
Note also that the inverse $\beta$ of tunneling distance 
in the expression of photoluminescence decay 
scales by $\alpha$, {\it i.e. }, $\alpha \beta$.
In other words, 
the effective tunneling distance increases 
when $\alpha$ is decreased by energetic disorder . 
In Fig. 3, theoretical lines obtained from the inverse Laplace transform of 
$\hat{R} \left(s \right) = 1 - s \hat{N} \left( s \right)$, where $\hat{N} \left( s \right)$ is given by Eq. (\ref{seriesrate}), 
are presented together with the measured photoluminescence decay at $145$ [K]. 
\cite{FDAMurayama,SekiPRB} 
$\gamma_{\rm r}$ is varied while keeping  
$\left( \gamma_{\rm r}^{3/2} /k_0 \right)^2 \sim 6.8 \times 10^{13}  \mbox{[s}^{-1}\mbox{]}$. 
This value of $\left( \gamma_{\rm r}^{3/2} /k_0 \right)^2$ 
guarantees the amplitude of asymptotic power law decay at other temperatures 
as explained above. 
The long time asymptote is well approximated by power law decay, $t^{- \delta}$ 
with $\delta = 2/\alpha +1$, where $\alpha \sim 0.6$ at $145$ [K] is obtained from 
Eq. (\ref{alphaT}). 
The asymptotic time region described by power law $t^{- \delta}$ increases with increasing 
$\gamma_{\rm r}$. 
In the case of $\gamma_{\rm r} = 10^{12} \mbox{[s}^{-1}\mbox{]}$, 
the line in Fig. 3 is described by power law decay in the entire time region of the Figure. 
Experimental data are close to the line of 
 $\gamma_{\rm r} = 10^{8} \mbox{[s}^{-1}\mbox{]}$. 
 In the case of  $\gamma_{\rm r} = 10^{6} \mbox{[s}^{-1}\mbox{]}$, 
 the onset time of asymptotic power law decay 
appears later than that of
experimental data. 
Before  the onset time,  
 the slope increases with increasing the delay time  after 
 the light pulse. 
 The value of $\gamma_{\rm r} = 10^8 \mbox{[s}^{-1}\mbox{]}$ is smaller than the values 
 $\gamma_{\rm r} \sim 10^{12}-10^{15} \mbox{[s}^{-1}\mbox{]}$ estimated from the time-of-flight technique. \cite{Street,MurayamaMC5,MurayamaJJAP}
In Fig. 4 $\gamma_{\rm r}$ is fixed to the value of $10^{12} \mbox{[s}^{-1}\mbox{]}$ and the value of $k_0$ is varied. 
The relation $\left( \gamma_{\rm r}^{3/2} /k_0 \right)^2 \sim 6.8 \times 10^{13} \mbox{[s}^{-1}\mbox{]}$ 
holds for $k_0 \sim 10^{11} \mbox{[s}^{-1}\mbox{]}$. 
For other values of $k_0$, 
the temperature dependence of photoluminescence intensity at 
$10\mbox{[}\mu\mbox{s]}$ is not reproduced. 
Nevertheless, we look for the possibility of fitting experimental values of photo-luminescence 
at $145$ [K], by assuming $\gamma_{\rm r} = 10^{12} \mbox{[s}^{-1}\mbox{]}$. 
In the case $k_0$ is smaller than $\gamma_{\rm r}$, namely, 
$\frac{1}{2 \alpha \beta} \sqrt{\frac{k_0^\alpha}{D_\alpha}} \leq1$, 
the photoluminescence decay is described by power law, $t^{- \delta}$ with the exponent 
$\delta = \alpha/2 + 1 \sim 1.3$ in the entire time region of the Figure. 
In the case of $k_0 > \gamma_{\rm r}$, 
namely, $\frac{1}{2 \alpha \beta} \sqrt{\frac{k_0^\alpha}{D_\alpha}} > 1$, 
the photoluminescence decay is described by power law $t^{-1.6}$ with the exponent 
larger than the value, $ \alpha/2 + 1 \sim 1.3$. 
Since the experimental data deviate from the power law $t^{-\alpha/2+1}$ 
at early times, 
they are not reproduced at early times by theoretical lines with 
$\gamma_{\rm r} = 10^{12}   \mbox{[s}^{-1}\mbox{]}$. 
Only the asymptotic decay is reproduced 
in the case of $\frac{1}{2 \alpha \beta} \sqrt{\frac{k_0^\alpha}{D_\alpha}} \leq 1$. 
When $k_0$ is large,  namely, 
$\frac{1}{2 \alpha \beta} \sqrt{\frac{k_0^\alpha}{D_\alpha}} > 1$, 
the photoluminescence decay is not described by 
the algebraic decay expressed by 
Eq. (\ref{powerPL}) in the time range described in Fig. 4. 
Nevertheless, 
the decay is approximately described by the power law of time 
with the exponent larger than $\alpha/2 +1$. 
In this limit, 
not only hopping but also 
the distance dependence of the intrinsic rate  
controls the main part of the photoluminescence decay.


\setcounter{equation}{0}
\section{Discussion and conclusions}
\vspace{0.5cm}
\label{section:Discussion}

We have derived methods to calculate reaction rates and survival probabilities for  
a continuous time random walk model with long-range reactivity. 
In Eq. (\ref{rrate})-(\ref{subprob}), 
survival probabilities and recombination rates are expressed in terms of the probability of just arriving at 
$r$ in the time interval between $t$ and $t+ d\, t$, 
which is obtained by the inverse Laplace transform of the solution of Eq. (\ref{continuouseta}). 
We also present more convenient alternative method, 
Eq. (\ref{raterho}) together with Eq. (\ref{contrho}). 
In this method, 
recombination rates and survival probabilities under dispersive diffusion are derived from those under normal diffusion with $\alpha=1$, 
by substituting 
$D_1 \rightarrow s^{1-\alpha} D_\alpha$ and $k_1 \left(\vecr \right) \rightarrow s^{1-\alpha} k_\alpha \left(\vecr \right)$
in the Laplace domain.  
The effective reactivity is related to the intrinsic reaction rate by Eq. (\ref{defintrinsicrate}).
The effective reactivity in the case of tunneling, 
$\gamma_{\rm rc} \left( \vecr \right) = k_0 e^{-2 \beta r}$, 
under the dispersive diffusion 
is expressed as $k_\alpha \left( r \right) = k_0^\alpha \exp \left( - 2 \alpha \beta r \right)$. 
Therefore, the effective tunneling distance $\left(\alpha \beta \right)^{-1}$ increases 
when $\alpha$ is decreased 
by energetic disorder. 
In the case of energy transfer, $\gamma_{\rm rc} \left( \vecr \right) = k_0 (R_F/r)^6$, 
with F\"{o}rster radius $R_F$, 
the effective reactivity is given by 
$k_\alpha \left( \vecr \right) = k_0^\alpha (R_F/r)^{6\alpha}$.  
The exponent $6\alpha$ is decreased with decreasing $\alpha$ 
and the long-range nature of energy transfer is enhanced.

In order to confirm the validity of the methods, 
we study the square well sink model under dispersive diffusion. 
\cite{Teramoto,Wilemski,Doi,Tachiya}
The escape probability and the Laplace transform of survival probability are known for normal diffusion. \cite{Tachiya}
Therefore, these quantities under dispersive diffusion are obtained by introducing 
substitution in the Laplace domain, 
$D_1 \rightarrow s^{1-\alpha} D_\alpha$ and $k_0 \rightarrow s^{1-\alpha} k_0^\alpha$. 
The results are compared to direct numerical simulations of random walk and excellent agreement is found 
between them.

Finally, the method is applied to obtain recombination rates by tunneling under dispersive diffusion. 
In the previous work, 
the relation $\delta = \alpha/2 + 1$ is confirmed for the experimental data of a-Si:H 
and the scaling relation of the amplitude, 
$R(t) \sim \frac{1}{C^\alpha t^{\alpha/2+1}}$ with 
$C^2 \sim 6.8 \times 10^{13} \mbox{[s}^{-1}\mbox{]}$ is obtained. \cite{SekiPRB}
We found that these relations are observable in the case 
$k_0$ is small compared to the rate of escape of an electron 
from the region within the distance 
$1/\left(\alpha \beta \right) \sim 10-30 \mbox{\AA}$ 
from a hole, namely, 
$\frac{1}{2 \alpha \beta} \sqrt{\frac{k_0^\alpha}{D_\alpha}} < 1$. 
The experimental data is close to the theoretical line with 
$\gamma_{\rm r} = 10^8 \mbox{[s}^{-1}\mbox{]}$ 
over the whole time range measured. 
However, 
the value of $\gamma_{\rm r} = 10^8 \mbox{[s}^{-1}\mbox{]}$ is smaller than the values 
 $\gamma_{\rm r} \sim 10^{12}-10^{15} \mbox{[s}^{-1}\mbox{]}$ estimated from the time-of-flight technique. \cite{Street,MurayamaMC5,MurayamaJJAP}
The discrepancy could be due to some unknown factors which may change the recombination rate such 
 as the initial distribution of an electron around a hole, electrostatic interaction energy between them. 
In the presence of distributed site energies of localized states, 
Coulombic interaction can be ignored in the case  
the distance between a hole and an electron is larger than 
the characteristic length $R_{\rm s} \equiv \sqrt{e^2 b/ \left(\epsilon k_{\rm B} T_0 \right)} 
\sim 40 \mbox{\AA}$, 
where $\epsilon \sim 10$ is the dielectric constant of a-Si:H. \cite{SekiPRB}
This distance is relatively small but still it is comparable to the hopping distance. 
It should be also noticed that the hopping frequency estimated from the time-of-flight technique 
is also approximate. 
Since the correct value of the hopping frequency is not known,  
it is not meaningful to introduce 
the distribution of initial positions of electrons which is not known either. 
When $k_0$ is large,  namely, 
$\frac{1}{2 \alpha \beta} \sqrt{\frac{k_0^\alpha}{D_\alpha}} > 1$, 
the photoluminescence decay is 
approximately described by the power law of time 
with the exponent larger than $\alpha/2 +1$ as shown in Fig. 4.  
In this limit, 
not only hopping but also 
the distance dependence of the intrinsic rate  
controls the main part of the photoluminescence decay.

\acknowledgments

This work is supported by the COE development program of 
the Ministry of Education, Culture, Sports, Science and Technology(MEXT) .

\appendix
\section{Relation between $\hat{\rho} \left( \vecr, s \right)$ and the Laplace transform of  the real density}
\label{appA}

By using the first equalities of Eqs. (\ref{Laplacepsiout}) and (\ref{Laplacepsireact}), we have, 
\begin{eqnarray}
 1-\hatpsiout  \left( \vecr, s \right)  -\hatpsireact  \left( \vecr, s \right)
 &=& \frac{s }
{s+\gamma_{\rm rc} \left( \vecr \right) }
\left[ 1 - \hat{\psi} \left(s + \gamma_{\rm rc} \left( \vecr \right)  \right) \right] .
\end{eqnarray}
Substituting the second equality  of Eq. (\ref{Laplacepsiout}) into the above equation yields, 
\begin{eqnarray}
\frac{\left[ 1-\hatpsiout  \left( \vecr, s \right)  -\hatpsireact  \left( \vecr, s \right) \right]}{s} &=& \frac{1}
{s+\gamma_{\rm rc} \left( \vecr\right) }  
\frac{\pi \alpha}{\sin \pi \alpha} \left( \frac{s+\gamma_{\rm rc} \left( \vecr \right) }{\gamma_{\rm r}} \right)^{\alpha} .
\label{relation1}
\end{eqnarray}
In the case of $\alpha=1$, 
the right hand side of Eq. (\ref{relation1}) is independent of $\gamma_{\rm rc} \left( \vecr \right)$
and can be shown to be identical to $\left[1-\hat{\psi} (s) \right]/s$ by  
Eq. (\ref{Laplacepsi}). 
Therefore, $\hat{\rho}  \left( \vecr, s \right) $ defined by 
$\hat{\eta}  \left( \vecr, s \right) \left[1-\hat{\psi} \left(  s \right) \right]/s$ coincides with 
the Laplace transformation of the 
real density given by 
$
\hat{\eta}    \left( \vecr, s \right) \left[ 1-\hatpsiout  \left( \vecr, s \right)  -\hatpsireact  \left( \vecr, s \right) \right] /s
$. 
However, they are generally different for $\alpha < 1$. 

\section{Derivation of Eq. (\ref{recombrateasym})}
\label{appB}
By substituting Eq. (\ref{seriesMdB}) into Eq. (\ref{seriesrate}), 
we obtain, 
\begin{eqnarray}
\hat{N} \left( s \right) 
&\sim& \frac{1}
{s \Gamma \left(1 +\frac{1}{\alpha \beta} \sqrt{\frac{s^\alpha}{D_\alpha}} \right)}
\frac{1}
{\displaystyle \sum_{n=0}^1 
\frac{\left( \frac{1}{2\alpha \beta} \sqrt{\frac{k_0^\alpha}{D_\alpha}}\right)^{2n}}
{n! \Gamma \left(n + 1 +\frac{1}{\alpha \beta} \sqrt{\frac{s^\alpha}{D_\alpha}} \right)} }
\nonumber \\
&=&  \frac{1}{s}
\frac{1}{
1 + 
\frac{
( \frac{1}{2\alpha \beta} 
\sqrt{k_0^\alpha / D_\alpha
} \,
)^2}
{
1+ \frac{1}{\alpha \beta}\sqrt{s^\alpha/D_\alpha}}
} .
\label{ap1}
\end{eqnarray}
In the limit of $s \rightarrow 0$, 
Eq. (\ref{ap1}) leads to 
\begin{eqnarray}
\hat{R} \left(s \right) =1 -s \hat{N} \left( s \right) 
\sim - \frac{\displaystyle \left( \frac{1}{2\alpha \beta} \sqrt{\frac{k_0^\alpha}{D_\alpha}}\right)^2}
{\displaystyle \left[ 1+ \left( \frac{1}{2\alpha \beta} \sqrt{\frac{k_0^\alpha}{D_\alpha}}\right)^2 \right]^2} 
\frac{1}{\alpha \beta}\sqrt{\frac{s^\alpha}{D_\alpha}} .
\label{ap2}
\end{eqnarray}
The inverse Laplace transform using Tauberian theorem  
yields Eq. (\ref{recombrateasym}). 
By expansion of Eq. (\ref{ap1}) in terms of series of $1/\sqrt{s^\alpha}$ and then performing the inverse Laplace transform, 
we obtain   
the survival probability, 
\begin{eqnarray}
N(t) \sim E_{\frac{\alpha}{2}} \left[ -\left(\alpha \beta \sqrt{D_\alpha} 
+ \frac{k_0^\alpha}{4 \alpha \beta \sqrt{D_\alpha}} \right) t^{\alpha/2} \right]
+ \alpha \beta \sqrt{D_\alpha t^\alpha} 
E_{\frac{\alpha}{2}, \frac{\alpha}{2}+1} \left[ -\left(\alpha \beta \sqrt{D_\alpha} 
+ \frac{k_0^\alpha}{4 \alpha \beta \sqrt{D_\alpha}} \right) t^{\alpha/2} \right], 
\nonumber \\
\end{eqnarray}
expressed 
in terms of 
the generalized Mittag-Leffler function defined by \cite{Podlubny}
\begin{eqnarray}
E_{a,b} \left( z \right) \equiv \sum_{k=0}^\infty \frac{\displaystyle z^k}{\displaystyle \Gamma \left(a k + b \right)} ,
\label{gML}
\end{eqnarray}
where $E_a \left( z \right) \equiv E_{a,1} \left( z \right) $. 
In the same way the recombination rate is obtained as 
\begin{eqnarray}
R(t) &\sim& 1- \frac{1}{t} E_{\frac{\alpha}{2},0} \left[ -\left(\alpha \beta \sqrt{D_\alpha} 
+ \frac{k_0^\alpha}{4 \alpha \beta \sqrt{D_\alpha}} \right) t^{\alpha/2} \right] 
\nonumber \\
& & 
- \frac{\alpha \beta \sqrt{D_\alpha t^\alpha} }{t}
E_{\frac{\alpha}{2}, \frac{\alpha}{2}} \left[ -\left(\alpha \beta \sqrt{D_\alpha} 
+ \frac{k_0^\alpha}{4 \alpha \beta \sqrt{D_\alpha}} \right) t^{\alpha/2} \right] .
\end{eqnarray}
We can confirm that the above equation reduces to 
Eq. (\ref{recombrateasym}) by 
the asymptotic expansion of 
the generalized Mittag-Leffler function,  \cite{Podlubny}
\begin{eqnarray}
E_{a,b} \left( z \right) \sim - \sum_{k=1} \frac{\displaystyle z^{-k}}{\displaystyle \Gamma \left(b-a k \right)} .
\label{gMLasymd}
\end{eqnarray}

In the case of normal diffusion, 
the inverse Laplace transform of Eq. (\ref{ap1}) becomes  
\begin{eqnarray}
N(t) \sim \frac{1}{1+ \frac{k_0}{4 \beta^2 D_1}} 
\left\{ 1 + \frac{k_0}{4 \beta^2 D_1} 
\exp \left[ \left( \beta \sqrt{D_1} 
+ \frac{k_0}{4 \beta \sqrt{D_1}} \right)^2 t \right]
\mbox{erfc} \left[ \left( \beta \sqrt{D_1} 
+ \frac{k_0}{4 \beta \sqrt{D_1}} \right) \sqrt{t} \right] \right\} ,
\nonumber \\
\end{eqnarray}
where $\mbox{erfc} (z) \equiv \frac{2}{\sqrt{\pi}} \int_z^\infty d\, t \exp \left( - t^2 \right)$. 
\cite{Abramowitz}

In this paper, 
we focused our attention on the limit of $r_0 = R$. 
As a reference, we list a result for $r_0 \geq R$. 
In the case of  
$R \rightarrow 0$, 
Eq. (\ref{lalpacetd}) is rewritten for an arbitrary initial position $r_0$ as, 
\begin{eqnarray}
\hat{\rho} \left( r_>, r_<, s \right) =
\frac{s^{\alpha-1}}{4\pi D_\alpha \alpha \beta} 
\left(  \hat{q} \left( r_<, s \right) - 
\frac{\displaystyle K_{\frac{1}{\alpha \beta}\sqrt{\frac{s^\alpha}{D_\alpha}}} 
\left( \frac{1}{\alpha \beta} \sqrt{\frac{k_0^\alpha}{D_\alpha}} \right)}
{\displaystyle I_{\frac{1}{\alpha \beta}\sqrt{\frac{s^\alpha}{D_\alpha}}} 
\left( \frac{1}{\alpha \beta} \sqrt{\frac{k_0^\alpha}{D_\alpha}} \right)}  \hat{p} \left( r_<, s \right) \right) 
 \hat{p} \left( r_>, s \right) , 
\end{eqnarray}
where $r_>= \max \left(r, r_0 \right)$ and $r_<= \min \left(r, r_0 \right)$.



\newpage
\vskip1cm
\noindent
{\bf Fig. 1:}~
The escape probability versus $D_{\alpha}/\left( R^2 k_0^\alpha \right)$.
The initial distance is $r_0=2R$ and $b/R=0.1$.
The symbols denote simulation data. 
The circles, triangles and 
crosses correspond to the 
$\alpha$ values of $0.5$, $0.75$ and $1.0$, respectively.
The line shows the analytical result of Eq. (\ref{escWF}).

\vskip1cm
\noindent
{\bf Fig. 2:}~
The survival probability as a function of time for $\alpha=0.5$.
$t$ is normalized by 
$\gamma_{\rm r}$, $r_0/R=2$ and $b/R=0.1$.
$k_0 \left( R^2/ D_\alpha \right)^{1/\alpha} =10^0,10^1,10^2$ from top to bottom.
Solid lines are the numerical Laplace inversion of Eq. (\ref{suvWF}). 
Dots represent the results of numerical simulations.

\vskip1cm
\noindent
{\bf Fig. 3:}~
Decay of photoluminescence in a-Si:H  excited with the pulsed laser of photon energy of $2.32$ [eV]. 
Dots show the observed decay in the whole time range measured. 
The photoluminescence has been observed at 145[K]. \cite{FDAMurayama,SekiPRB} 
Lines are calculated from the inverse Laplace transform of 
$\hat{R} \left(s\right) = 1 - s \hat{N}(s)$, where $\hat{N} \left( s \right)$ is given by Eq. (\ref{seriesrate}) with 
 $\left( \gamma_{\rm r}^{3/2} /k_0 \right)^2 \sim 6.8 \times 10^{13}  \mbox{[s}^{-1}\mbox{]}$ being fixed,  
$\alpha = 0.15 + 145/313.4 \sim 0.61$, $b=10\mbox{\AA}$ and $\beta=0.05 \mbox{[1/\AA]}$. 
Amplitudes are adjusted by comparison to experimental data. 
The short dashed line, solid line and long dashed line represent $\gamma_{\rm r} = 10^6  \mbox{[s}^{-1}\mbox{]}$, 
$\gamma_{\rm r} = 10^8  \mbox{[s}^{-1}\mbox{]}$ and $\gamma_{\rm r} = 10^{10}  \mbox{[s}^{-1}\mbox{]}$, respectively.

\vskip1cm
\noindent
{\bf Fig. 4:}~
Decay of photoluminescence in a-Si:H  excited with the pulsed laser of photon energy of $2.32$ [eV]. 
Dots show the observed decay in the whole time range measured. 
The photoluminescence has been observed at 145[K]. \cite{FDAMurayama,SekiPRB} 
Lines are calculated from the inverse Laplace transform of 
$\hat{R} \left(s\right) = 1 - s \hat{N}(s)$, where $\hat{N} \left( s \right)$ is given by Eq. (\ref{seriesrate}) 
with  $\gamma_{\rm r}= 10^{12}  \mbox{[s}^{-1}\mbox{]}$,  
$\alpha = 0.15 + 145/313.4 \sim 0.61$, $b=10\mbox{\AA}$ and $\beta=0.05 \mbox{[1/\AA]}$. 
Amplitudes are adjusted by comparison to experimental data. 
The thin solid line, thick solid line and short dashed line represent $k_0 = 10^6  \mbox{[s}^{-1}\mbox{]}$, 
$k_0 = 10^{11} \mbox{[s}^{-1}\mbox{]}$ and $k_0 = 10^{12} \mbox{[s}^{-1}\mbox{]}$, respectively. 
They obey almost the same power law, $1/t^\delta$, with  the exponent $\delta = \alpha/2 + 1 \sim 1.3$. 
The dashed-and-dotted line and long dashed line represent 
$k_0 = 10^{13} \mbox{[s}^{-1}\mbox{]}$ and $k_0 = 10^{14} \mbox{[s}^{-1}\mbox{]}$, respectively.
They are well approximated by the power law, $\sim 1/t^{1.6}$, with the exponent $1.6$ larger than  
$\delta = \alpha/2 + 1 \sim 1.3$.

\newpage
\begin{flushright}
Fig. 1, K. Seki, M. Wojcik, and M. Tachiya
\end{flushright}
\vspace{5cm}
\mbox{ }\\
\begin{figure}[h]
\includegraphics[width=7cm]{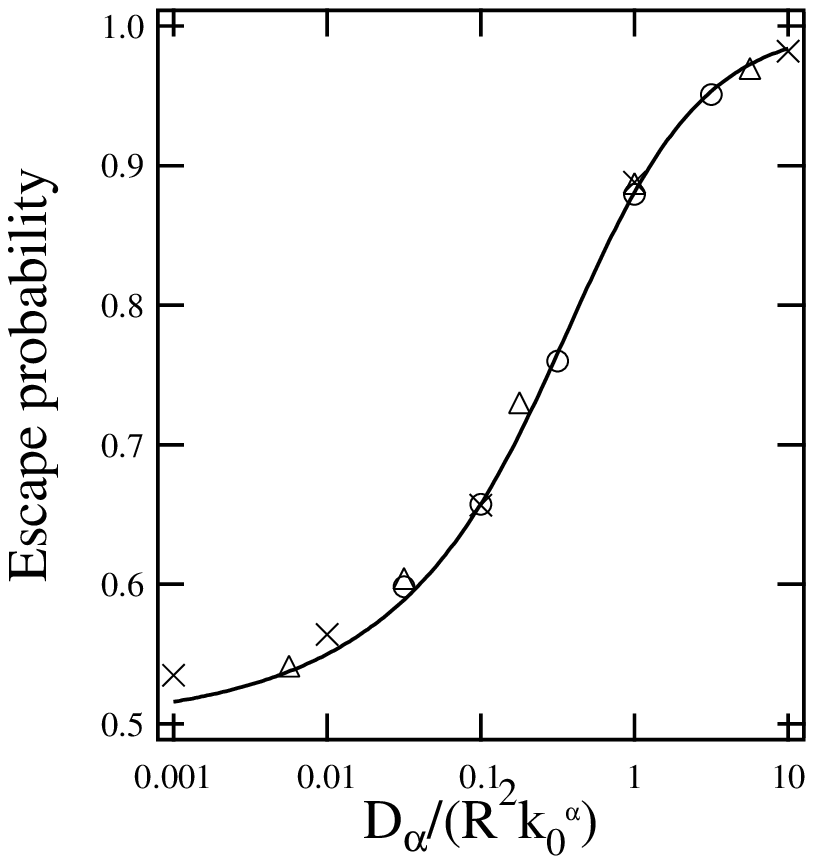}
\end{figure}

\newpage
\begin{flushright}
Fig. 2, K. Seki, M. Wojcik, and M. Tachiya
\end{flushright}
\vspace{5cm}
\mbox{ }\\
\begin{figure}[h]
\includegraphics[width=7cm]{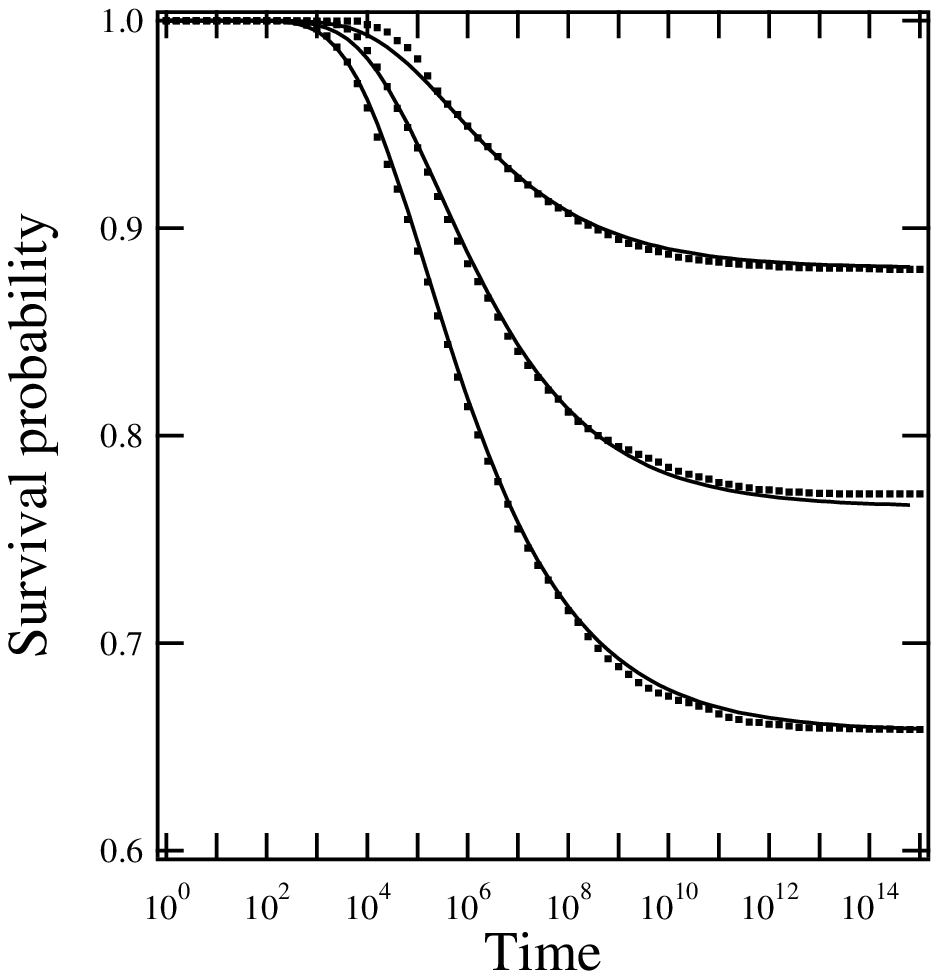}
\end{figure}

\newpage
\begin{flushright}
Fig. 3, K. Seki, M. Wojcik, and M. Tachiya
\end{flushright}
\vspace{5cm}
\mbox{ }\\
\begin{figure}[h]
\includegraphics[width=7cm]{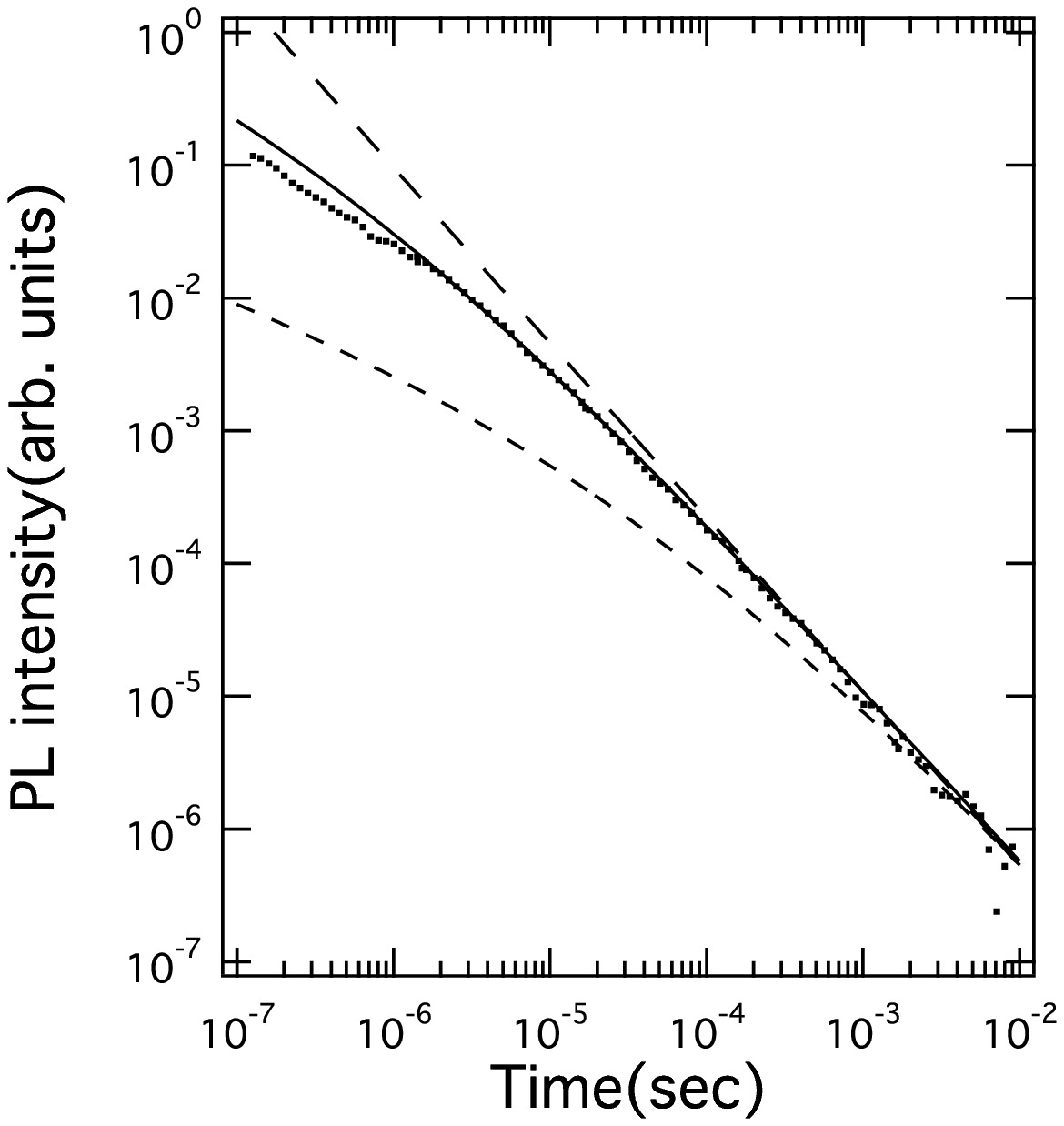}
\end{figure}

\newpage
\begin{flushright}
Fig. 4, K. Seki, M. Wojcik, and M. Tachiya
\end{flushright}
\vspace{5cm}
\mbox{ }\\
\begin{figure}[h]
\includegraphics[width=7cm]{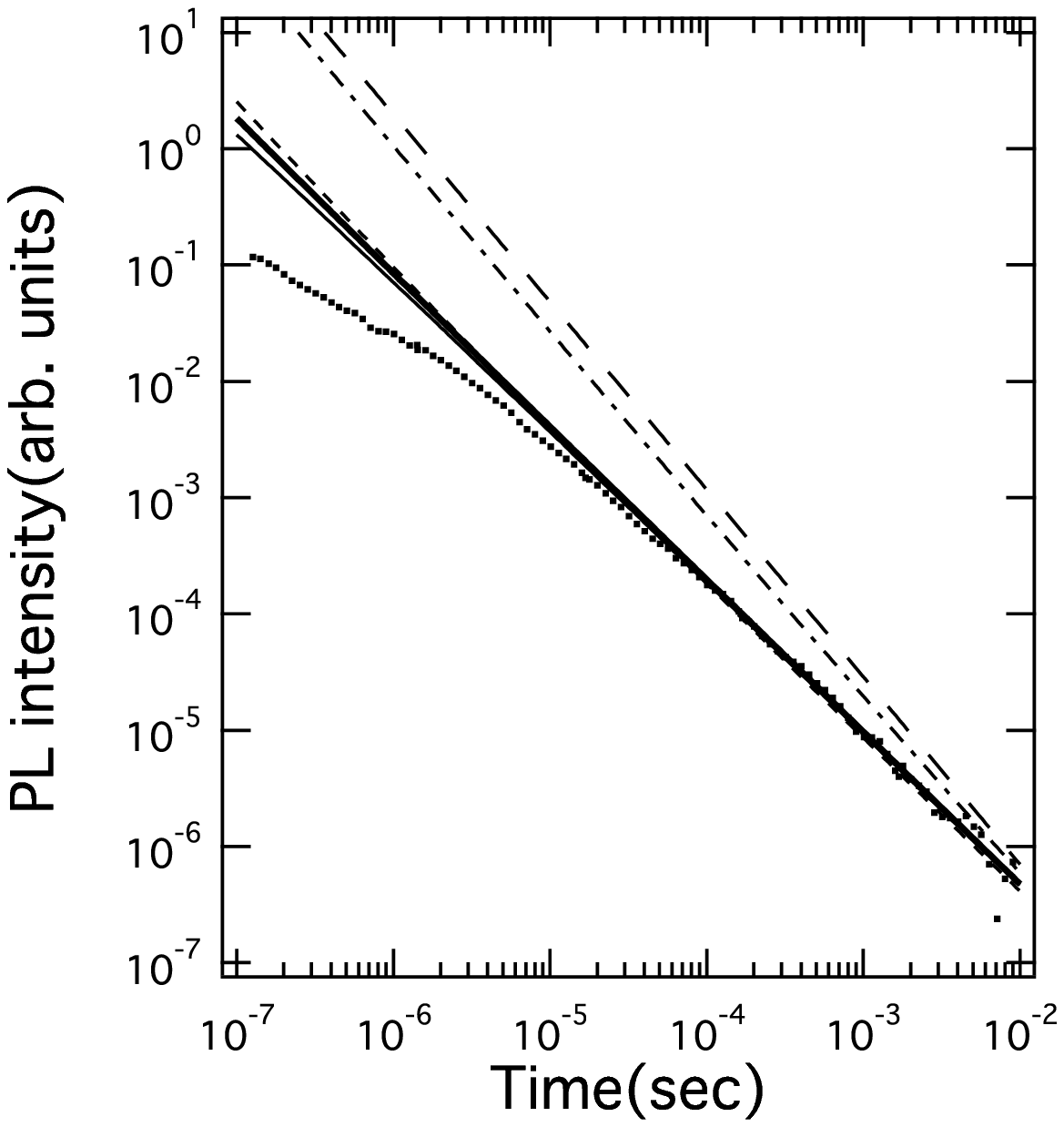}
\end{figure}

\end{document}